\documentclass[prb,amsmath,amssymb,twocolumn]{revtex4-1} 
\usepackage{graphicx}
\usepackage[table]{xcolor}
\usepackage[english]{babel}

\begin{document}
          
\title{Spectral Fingerprints of Electron-Plasmon Coupling}
\author{Fabio Caruso}
\author{Feliciano Giustino}
\affiliation{Department of Materials, University of Oxford, Parks Road, Oxford OX1 3PH, United Kingdom}
\date{\today}
\date{\today}

\begin{abstract}
We investigate the spectroscopic fingerprints of electron-plasmon coupling
in integrated (PES) and angle-resolved photoemission spectroscopy (ARPES).
To account for electron-plasmon interactions at a reduced computational cost, 
we derived the plasmonic polaron model, an approach based on the 
cumulant expansion of many-body perturbation theory that circumvents 
the calculation of the $GW$ self-energy.
Through the plasmonic polaron model, we predict the complete spectral functions 
and the effects of electron-plasmon coupling for Si, Ge, GaAs, and diamond. 
Si, Ge, and GaAs exhibit well defined plasmonic polaron band structures, i.e.,
broadened replica of the valence bands red-shifted by the plasmon energy.  
Based on these results, (i) we assign the structures of the plasmon satellite 
of silicon (as revealed by PES experiments) to plasmonic Van Hove singularities occurring 
at the $L$, $\Omega$, and $X$ high-symmetry points and (ii)
we predict the ARPES signatures of electron-plasmon coupling for Si, Ge, GaAs, and diamond.
Overall, the concept of plasmonic polaron bands emerges as a new paradigm for the interpretation of
electronic processes in condensed matter, and the theoretical approach presented here 
provides a computationally affordable tool to explore its effect in a broad set of materials.
\end{abstract}

\maketitle   

\section{Introduction}
The coupling of electrons and bosonic excitations, such as phonons, magnons, and  collective charge-density fluctuations ({\it plasmons}), 
may significantly alter the electronic properties of solids, triggering
emergent phenomena which may not be explained within the context of 
a simple single-particle picture. 
A prototypical example is the formation of Cooper pairs,\cite{Cooper1956}
resulting from electron-phonon coupling, and the emergence of superconductivity.\cite{BCS1957}
Whereas phonons have a characteristic energy in the range of $10-100$~meV,
the energy of {\it plasmons} is typically between $1$ and $10$~eV. 
The spectral signatures of electron-plasmon coupling, thus, occur at energies 
comparable to those of the ordinary quasiparticle states which are easily accessible 
in integrated (PES) and angle-resolved photoemission spectroscopy (ARPES).

Theoretical approaches for the description of electron-plasmon interactions from first principles are 
generally based on many-body perturbation theory by combining the one-shot $GW$ approximation\cite{PhysRev.139.A796,Hybertsen1986} with the cumulant expansion 
approach\cite{Lundqvist,PhysRevB.1.471} ($GW$+C).
The cumulant expansion,  formally exact for the case of an isolated core-electron,\cite{PhysRevB.1.471}
stems from the solution of an electron-boson coupling Hamiltonian analogous to the case of electron-phonon coupling. 
The applicability of this method has been extended beyond the case of core satellites and its has proven useful 
in the description of quasiparticle dynamics,\cite{Gumhalter2012163,PhysRevB.72.165406} and 
valence plasmon satellites in the {\it integrated} 
photoemission spectra of the homogeneous electron gas,\cite{Holm1997,Kas2014} 
simple metals,\cite{PhysRevLett.77.2268} silicon,\cite{Guzzo2012} and graphene.\cite{Guzzo2014,PhysRevLett.110.146801}

First-principles calculations based on the 
$GW$+cumulant approach have recently unveiled  
novel spectroscopic signatures of electron-plasmon coupling in the 
angle-resolved photoemission spectra (ARPES) of solids: 
the {\it band structures of plasmonic polarons}.\cite{Caruso2015}
Plasmonic polarons are elementary excitations resulting from 
the simultaneous excitation of an electron, and a plasmon and lead to the formation of 
dispersive spectral features in systems characterized by well defined 
plasmonic excitations (as, e.g., silicon, and transition-metal dichalcogenides) which follow closely the 
dispersive character of the ordinary quasiparticle band structure.  
In particular, the dispersive nature of plasmonic polarons leads to the  
formation of well-defined plasmonic polaron bands which manifest
themselves as broadened replica of the ordinary quasiparticle 
band structure, red-shifted by the plasmon energy.
The existence of plasmonic polaron bands has recently been {\it confirmed} by 
ARPES measurements 
of the full band structure of silicon in an energy window 
that extends up to 35~eV below the Fermi energy.\cite{Lischner2015}

In this work, we introduce the {\it plasmonic polaron model}, 
an approach for the computation of plasmonic polaron band structures that  
circumvents the calculations of the $GW$ self-energy.
Beside illustrating analytically the emergence of plasmonic polaron bands within the 
$GW$+cumulant formalism, the plasmonic polaron models allows to account for the effects 
of electron-plasmon coupling in the spectral properties of solids at a reduced numerical cost. 
We thus discuss electron-plasmon interactions in Si, GaAs, Ge, and diamond. 
Our results reveal that the different substructures of the plasmon satellite of silicon, 
as measured by XPS, may be assigned to Van Hove singularities arising from the plasmonic polaron 
bands taking place at the $L$, $\Omega$, and $X$ high-symmetry points of the first Brillouin zone. 
For GaAs, Ge, and diamond, we predict plasmonic polaron bands in qualitative agreement with silicon, 
and we identify the spectral features that may guide the observation of plasmonic polaron 
bands in these compounds.
Overall our results indicate that plasmonic polaron bands provide a novel 
concept that may prove generally applicable in the interpretation of the 
integrated and angle-resolved spectral 
properties of solids. 

\section{Plasmonic Polaron Model}\label{sec:model}

In the following, we derive 
an approach to predict the band structures of plasmonic polarons at the numerical cost 
of a band structure calculation, circumventing the explicit calculation of the $GW$ self-energy.
The cumulant expansion for the single-particle Green function\cite{PhysRevB.1.471,1402-4896-21-3-4-039,PhysRevLett.77.2268,Holm1997, PhysRevLett.107.166401, PhysRevLett.110.146801,Guzzo2012,Kas2014,Guzzo2014} 
provides an ideal framework for the description of electron-plasmon interactions in solids. 
In particular, the starting point for the following discussion is provided by 
the derivation of the  $GW$+cumulant approach reported in Ref.~\onlinecite{PhysRevLett.77.2268}
(referred to as $GW+C_{\rm AHK}$ hereafter).
If one neglects rare events in which multiple plasmons are emitted 
simultaneously with the excitation of a photo-hole, 
the $GW$+C$_{\rm AHK}$ spectral function can 
be expressed as:\cite{PhysRevLett.77.2268}
  \begin{equation}\label{eq-spectrum1}
  A({\bf k},\omega) = \sum_{n} [ A_n^{\rm QP}({\bf k},\omega) + 
  A_n^{\rm QP}({\bf k},\omega)\ast A_n^{\rm C}({\bf k},\omega)]. 
  \end{equation}
The first term in Eq.~(\ref{eq-spectrum1}) accounts for the spectral signatures  
of quasiparticle excitations (i.e., in absence plasmon emission) 
and is defined as:
  \begin{equation}\label{eq.specfun}
  A_n^{\rm QP}({\bf k},\omega)  
  = \frac{1}{\pi} \frac{\Gamma_{n{\bf k}} (\omega)}
  {[\omega - \varepsilon_{n{\bf k}} -\Delta \Sigma_{n{\bf k}}(\varepsilon_{n{\bf k}})]^2 
  + [\Gamma_{n{\bf k}}(\omega)]^2},
  \end{equation}
where $\Delta\Sigma_{n{\bf k}}$ denotes the $G_0W_0$ quasiparticle 
correction\cite{PhysRev.139.A796,Hybertsen1986} to the 
Kohn-Sham eigenvalues\cite{PhysRev.136.B864,PhysRev.140.A1133} $\varepsilon_{n{\bf k}}$, 
and $\tau_{n{\bf k}}\equiv 1/\Gamma_{n{\bf k}}$ the quasiparticle lifetime.
Here and in the following we adopt Hartree atomic units.
As discussed in Ref.~\onlinecite{Caruso2015},
the convolution product in the second term of Eq.~(\ref{eq-spectrum1}) introduces novel dispersive features 
in the spectral function, which
account for events in which a photo-hole and a plasmon are excited simultaneously. 
The term $A_n^{\rm C}$ is defined as:\cite{PhysRevLett.77.2268} 
  \begin{equation}\label{eq-spectrum2}
  A_n^{\rm C}({\bf k},\omega)  
  =  \frac{\beta_{n{\bf k}}(\omega) - \!\beta_{n{\bf k}}(\varepsilon_{n{\bf k}}) - 
  \!(\omega-\varepsilon_{n{\bf k}})\!\left. \displaystyle\frac{\partial \beta_{n{\bf k}}}{\partial \omega}
    \right|_{\varepsilon_{n{\bf k}}}}
  {(\omega-\varepsilon_{n{\bf k}})^2},
  \end{equation}
where $\beta_{n{\bf k}}(\omega) = \pi^{-1}{\rm Im}\Sigma_{n{\bf k}}(\varepsilon_{n{\bf k}}-\omega)
\theta(\mu-\omega)$. 

  \begin{figure}
  \includegraphics[width=0.35\textwidth]{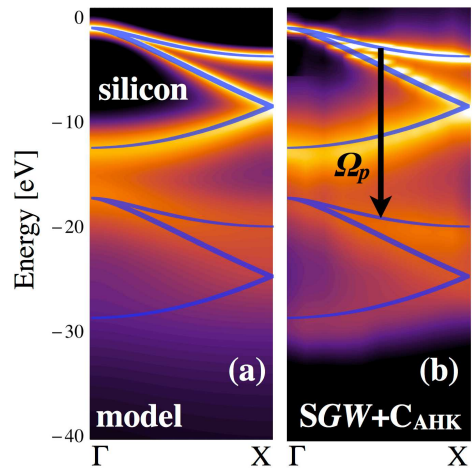}
  \caption{Angle-resolved spectral function of Silicon for momenta along the $\Gamma$-$X$ direction evaluated
  from (a) the plasmonic polaron model [Eq.~(\ref{eq-cumulant-appr1})]
  and (b) converged S$GW+C_{\rm AHK}$ calculations from Ref.~\onlinecite{Caruso2015}. 
  The vertical arrow indicates the plasmon energy $\Omega_p$ of silicon. 
  The {\it bare} PBE band structure and its shifted plasmonic polaron replica are shown in blue.
  }
  \label{fig:sgw-vs-model}
  \end{figure}
  \begin{figure*}
  \includegraphics[width=0.75\textwidth]{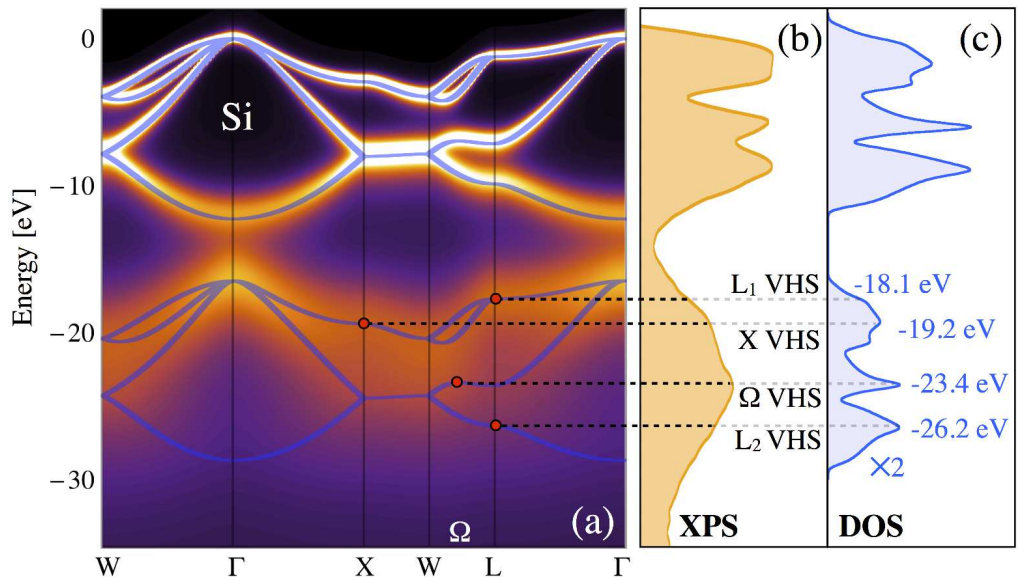}
  \caption{(a) Spectral function of silicon determined from the plasmonic polaron model [Eq.~(\ref{eq-cumulant-appr1})]. 
  The plasmonic polaron intensity has been magnified by a factor of 3 to enhance visibility.
  (b) Experimental (integrated) x-ray photoemission spectrum (XPS) of silicon 
  reproduced from Ref.~\onlinecite{PhysRevLett.107.166401}.
  (c) Quasiparticle and plasmonic polaron (magnified by a factor of 2) density of states (DOS). 
  Different peaks and shoulders of the plasmonic polaron resonance in the XPS spectrum are 
  attributed to plasmonic Van Hove singularities occurring on 
  different high-symmetry lines in the Brillouin zone (red circles). 
  }
  \label{fig:Sifull}
  \end{figure*}

To simplify the evaluation of Eq.~(\ref{eq-spectrum1}), we notice that 
the self-energy of systems characterized by well-defined plasmon resonances 
-- as, e.g., simple metals\cite{PhysRevLett.77.2268} (Al and Na) and semiconductors\cite{PhysRevB.81.115105}  (Si and GaAs) -- 
exhibits an imaginary part with a large plasmon peak at
$\omega = \varepsilon_{n{\bf k}} - \Omega_p$, where $\Omega_p$ is the plasmon energy. 
The plasmon peak in ${\rm Im}\Sigma$ stems directly from the plasmonic resonance of 
${\rm Im}\epsilon^{-1}(\omega)$, with $\epsilon$ being the 
dielectric function in the random-phase approximation (RPA). 
Therefore, the second term of Eq.~(\ref{eq-spectrum1}) may be simplified 
by assuming ${\rm Im}\Sigma_{n{\bf k}}\simeq \lambda_{n{\bf k}}L(\omega)$, where $ \lambda_{n{\bf k}}$ is 
a constant and $L$ 
a normalized Lorentzian function centered at $\omega=\varepsilon_{n{\bf k}} - \Omega_p$, that is:
$L(\omega)=({\gamma}/{\pi})/ [{(\omega-\varepsilon_{n{\bf k}} + \Omega_p)^2 + \gamma^2}]$.
The {\it ansatz} introduced above for ${\rm Im} \Sigma$ is an oversimplification 
of the structures of the self-energy that, as demonstrated below, suffices to reproduce 
the essential experimental spectral features arising from the electron-plasmon interaction. 

The Lorentzian {\it ansatz} for ${\rm Im} \Sigma$ can be derived within the framework 
of the ordinary plasmon-pole approximation\cite{Lundqvist2,Hybertsen1986,Godby1989,Engel1993,PhysRevB.76.195116,PhysRevB.84.241201,PhysRevB.88.125205,PhysRevLett.105.146401} for $GW$ calculations 
by considering non-dispersive bands and non-dispersive plasmons.

If the full-width at half maximum (FWHM) $\gamma$ is assumed to be sufficiently small ($\gamma \ll |\varepsilon_{n{\bf k}} - \Omega_p|$)
one may retain only the first term in the numerator of Eq.~(\ref{eq-spectrum2}), which simplifies to:
\begin{align}\label{eq-Ac-appr}
A^{\rm C}_n({\bf k, \omega}) = \frac{\beta_{n{\bf k}}(\omega)}{(\omega - \varepsilon_{n{\bf k}})^2}
\simeq\frac{\lambda_{n{\bf k}}}{\pi \Omega_p^2} L(\varepsilon_{n{\bf k}}-\omega). 
\end{align}
The convolution product in Eq.~(\ref{eq-spectrum1}) can be approximated using Eq.~(\ref{eq-Ac-appr}) and 
the spectral function reduces to:
\begin{align}\label{eq-cumulant-appr1}
 A ({\bf k},\omega) \simeq \sum_n \left[  A^{\rm QP}_n ({\bf k},\omega) + 
\frac{\lambda_{n{\bf k}}}{\pi\Omega_p^2}  
A^{\rm QP}_n ({\bf k},\omega+\Omega_p)  \right].
\end{align}
Equation~(\ref{eq-cumulant-appr1}) -- the central equation of the 
plasmonic polaron model --  illustrates
the emergence of band structure replicas as a result of electron-plasmon interactions. 
The resulting additional spectral features are the band structures of plasmonic polarons --
copies of the ordinary quasiparticle bands red-shifted by the plasmon energy $\Omega_p$.
These bands, which follow closely the momentum dependence of the quasiparticle energies, 
reveal the dispersive nature of plasmonic polarons.

In Eq.~(\ref{eq-cumulant-appr1}), 
the prefactor ${\lambda_{n{\bf k}}}/{\pi\Omega_p^2}$ determines the relative 
spectral weight of the plasmonic polaron features with respect to the quasiparticle peaks, and it 
may be related to the quasiparticle weight $Z_{n{\bf k}}$.  
In practice, the quasiparticle part of the spectral function is related to the quasiparticle weight through
$\int_{-\infty}^\mu A_{n{\bf k}}^{\rm QP}(\omega)d\omega = Z_{n{\bf k}}$.
Similarly, making use of Eq.~(\ref{eq-Ac-appr}) and of the normalization of $L(\omega)$ one can write:
\begin{align}
\int_{-\infty}^\mu &[ A_{n}^{\rm QP}({\bf k},\omega) \ast A_{n}^{\rm C}({\bf k},\omega)]d\omega 
= \frac{ \lambda_{n{\bf k}}} {\pi \Omega_p^2} Z_{n{\bf k}}.
\end{align}
Requiring the normalization of the spectral function
(that is, $\int_{-\infty}^\mu A_{n{\bf k}}(\omega) = 1$) it is possible to
express $\lambda_{n{\bf k}}$ explicitly in terms of $\Omega_p$
and $Z_{n{\bf k}}$:\footnote{The same expression can be derived combining the Lorentzian approximation for the 
self-energy with the expression $Z_{n{\bf k}}=\left[1+\int\frac{\beta_{n{\bf k}}(\omega)}{\omega^2}d\omega\right]^{-1}$ derived in 
Ref.~\onlinecite{0953-8984-11-42-201}.}
\begin{align}\label{eq-lambda}
\lambda_{n{\bf k}} = \pi \Omega_p^2 \frac{1-Z_{n{\bf k}}}{Z_{n{\bf k}}}. 
\end{align}

To further simplify the calculation of the plasmonic polaron band structures, 
we neglect the quasiparticle
correction to the Kohn-Sham eigenvalues (i.e., $\Delta\Sigma_{n{\bf k}} = 0$) in 
Eq.~(\ref{eq.specfun}),
and we introduce a simple analytical model for the 
quasiparticle lifetime.
According to Fermi liquid theory, 
the inverse lifetime of quasiparticle states close to the
Fermi energy increases quadratically with their energy , that is
$\tau_{n{\bf k}}^{-1}\propto (\varepsilon_{n{\bf k}} -\mu)^2$. 
Here we consider semiconductors, whereby ${\rm Im }\Sigma(\omega)=0$ 
for energies in the range $\omega\in[\mu-E_{\rm g},\mu]$, $E_{\rm g}$ being the quasiparticle band gap, owing to the absence of (electronic)
decay channels that may induce a de-excitation of the photohole. 
We combine these two ideas to define a simplified model for the electronic lifetimes of  semiconductors. 
Henceforth, we assume a quadratic dependence of the  
quasiparticle lifetimes on the frequency (relative to the Fermi energy), in the form:
$\Gamma(\omega)= \eta + \alpha (\omega + E_{\rm g})^2\theta(\mu-E_{\rm g}-\omega)$. 
In short, $\eta=0.05$~eV is introduced to avoid divergences close to the Fermi energy, whereas 
the term $E_{\rm g}$ ensures that the inverse 
lifetime of states in the energy window $\omega\in[\mu-E_{\rm g},\mu]$
is negligible due to the absence of decay channels (phonon-assisted decay processes are neglected here).
The parameter $\alpha$ is determined by imposing the equality between the integrated area of the broadening 
model defined above and the area underlying the imaginary part of the self-energy, which coincides with 
$\lambda_{n{\bf k}}$. In other words, we require
$\int_{\overline\omega}^\mu \Gamma(\omega)d\omega 
= \int_{\overline\omega}^\mu {\rm Im}\Sigma_{n{\bf k}} (\omega)d\omega= \lambda_{n{\bf k}}$,
where $\overline\omega$ denotes the energy such that ${\rm Im}\Sigma_{n{\bf k}} (\omega)\simeq0$ 
for $\omega<\overline\omega$
[in the following, $\overline\omega = 2(\varepsilon_{n{\bf k}}-\Omega_p)$]. 
Alternative broadening function models for $\Gamma$  may be employed, 
however the precise choice for $\Gamma$ does not alter the qualitative features 
of the plasmonic polaron band structures determined through Eq.~(\ref{eq-cumulant-appr1}). 

  \begin{figure*}
  \includegraphics[width=0.98\textwidth]{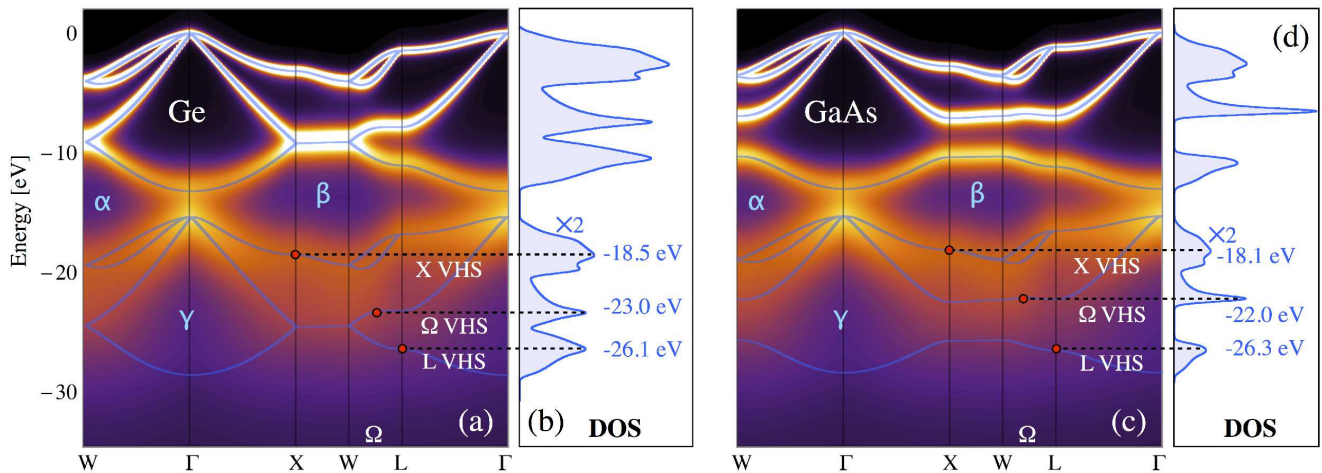}
  \caption{Spectral function of (a) Ge and (c) GaAs determined 
  from the plasmonic polaron model [Eq.~(\ref{eq-cumulant-appr1})].
  The plasmonic polaron intensity has been magnified by a factor of 3 to enhance visibility.
  $\alpha$, $\beta$, and $\gamma$ denote regions characterized by low density of states 
  due to the absence of plasmonic polarons and quasiparticle states.
  Panel (b) and (d): quasiparticle and plasmonic polaron DOS of Ge and GaAs.
  }
  \label{fig:Ge/GaAs}
  \end{figure*}

In summary, Eqs.~(\ref{eq.specfun}) and (\ref{eq-cumulant-appr1}) define a simple general procedure for predicting 
the plasmonic polaron band structures of systems characterized by well defined plasmonic excitations, 
{\it avoiding the explicit calculation of the $GW$ self-energy}.
The resulting model requires knowledge of a few parameters
which, for semiconductors,  may easily be inferred from experiment and/or 
first-principles calculations: 
the plasmon energy $\Omega_p$, the quasiparticle weight $Z_{n{\bf k}}$, and 
the quasiparticle band gap $E_{\rm g}$.
For $sp$-bonded semiconductors typically  $0.7<Z_{n{\bf k}}<0.9$.
In the following, we assume $Z_{n{\bf k}}\simeq 0.8$, as, e.g., for bulk Si and Ge.
To further reduce the number of external parameters, 
we obtain the plasmon energy $\Omega_p$ 
from a simple model for the frequency-dependent dielectric
function of semiconductors:\cite{Tosatti1971623}
$\Omega_p^2 = \omega_p^2+E_{\rm g}^2 $, 
where $\omega_p$ is the homogeneous electron gas plasma frequency 
at the valence electron density\cite{Tosatti1971623} and $E_{\rm g}$ the quasiparticle band gap. 
In the following, we consider the experimental value for the quasiparticle band gap $E_{\rm g}$.
Once the parameters $\Omega_p$, $E_{\rm g}$, and $Z$ are given, the evaluation of 
Eq.~(\ref{eq-cumulant-appr1}) can be performed through a simple band-structure calculation.
In the following, all calculations are performed in a plane-wave basis within the 
{\tt Quantum Espresso}
code.\cite{0953-8984-21-39-395502}$^,$\footnote{We employed the Perdew-Burke-Ernzerhof (PBE) 
version of the generalized-gradient approximation\cite{PhysRevLett.77.3865} for the
Kohn-Sham DFT exchange-correlation functional.\cite{PhysRev.136.B864,PhysRev.140.A1133}
Our atomistic models of the Si, Ge, GaAs, and diamond crystal structures are 
based on the experimental structural parameters. 
Only valence electrons are treated explicitly in our calculations, whereas 
core electrons are accounted for through 
Rappe-Rabe-Kaxiras-Joannopoulos ultrasoft pseudopotentials.\cite{PhysRevB.41.1227}
The momentum integrals over the first Brillouin zone are discretized over a 
$8\times8\times8$ homogeneous Monkhorst-Pack grid, and all plane waves 
up to a kinetic energy cutoff of 30~Ry are included in the calculation.} 

{ 
Before moving on to discuss electron-plasmon interactions in semiconductors, we briefly summarize 
the approximations employed in the plasmonic polaron model and we comment on their validity.
The only assumptions used until this point are that
(i) the imaginary part of the self-energy can be approximated by a sharp pole at the plasmon energy and 
(ii) the photo-emission process can be described within the {\it sudden approximation}. 

The assumption (i) is based on the fact that 
systems characterized by plasmonic excitations typically exhibit a well defined resonance in 
the imaginary part of the inverse dielectric function $\epsilon$  
(owing to its relation to the electron energy loss function\cite{PhysRev.113.1254}) which  introduces a corresponding 
plasmonic resonance in ${\rm Im}\Sigma$ via the relation ${\rm Im}\Sigma =  {\rm Im}[\epsilon^{-1}v] G$.
This approximation is thus justified for systems in which the 
plasmon-pole approximation yields a reasonable description of the inverse dielectric matrix.
The plasmon energy dispersion, neglected within approximation (i), would introduce an additional broadening 
of the self-energy but it is not expected to alter the qualitative spectral features obtained from the model. 
The approximation (ii) in practice corresponds to neglecting extrinsic effects to the 
photoemission process \cite{PhysRevLett.78.1528,Hedin1998,Rehr2000,Guzzo2012}.
Extrinsic effects, which account for the interactions between 
the photoelectron and the system after emission, are expected to introduce an 
additional broadening of the plasmonic polaron features and a renormalization of their 
intensity \cite{PhysRevLett.107.166401,Guzzo2012}. 
Additionally, in the following we estimate the 
plasmon energies from a simplified dielectric function model for semiconductors which assumes that 
collective charge density fluctuations involve the entirety of the valence
electrons and that electronic states are sufficiently delocalized\cite{pines1999elementary,mahan2000many} (conditions obeyed by 
$sp$-bonded systems as those considered here). 
For $d$- and $f$-electron systems, where these conditions may not apply, it would be more appropriate to employ 
plasmon energies derived either from experiment, from first-principles
calculations of the energy-loss function in the random-phase approximation, or from more elaborate dielectric function models.
%
} 

\section{Plasmonic Polarons and Plasmonic Van Hove Singularities}
As a first step, we validate the plasmonic polaron model 
by comparing the spectral function of silicon obtained from Eq.~(\ref{eq-cumulant-appr1})
with accurate first-principles calculations 
based on the Sternheimer$GW$+cumulant 
approach\cite{PhysRevB.81.115105,PhysRevB.88.075117,PhysRevLett.77.2268,Caruso2015} (S$GW+C_{\rm AHK}$).
In short, S$GW$ provides an accurate reference method for the 
calculation of the full spectral function. In particular, in S$GW$ 
(i) summations over empty states are avoided by computing the screened Coulomb interaction $W$ 
and Green's function through the iterative solution of the Sternheimer equation and 
(ii) beside the random-phase approximation (RPA) there are no further approximations 
(as, e.g., the plasmon-pole model) involved in the computation of the dielectric function. 
More details on the theoretical and numerical aspects underlying the 
Sternheimer-$GW$ approach can be found elsewhere.\cite{PhysRevB.81.115105,PhysRevB.88.075117,baroni01}
In Fig.~\ref{fig:sgw-vs-model}, we report the angle-resolved spectral function of Silicon
for momenta within the first Brillouin zone along the $\Gamma$-$X$ high-symmetry line as obtained 
from (a) the plasmonic polaron model [Eq.~(\ref{eq-cumulant-appr1})] 
and (b) the S$GW+C_{\rm AHK}$ approach. 
The PBE band structure of silicon and its plasmonic polaron replica (red-shifted by 
the plasmon energy $\Omega_p$) are superimposed as a continuous blue line for comparison.
Overall, as illustrated in Fig.~\ref{fig:sgw-vs-model}, the proposed plasmonic polaron 
model allows us to reproduce the qualitative features of S$GW+C_{\rm AHK}$ approach at 
the cost of a band-structure calculation. 
Based on this finding, we now move on to 
discuss the full spectral function of Si, GaAs, Ge, and diamond and the
spectral fingerprints of plasmonic polaron excitations in 
these compounds. 

%
In Fig.~\ref{fig:Sifull} we report (a) the full spectral function of silicon evaluated within the 
plasmonic polaron model and (c) the corresponding density of states (DOS).
The plasmonic polaron DOS has been calculated by ignoring the frequency 
dependent component of the broadening function ($\alpha=0$) to emphasize 
the structure arising from the different Van Hove singularities. 
{ To approximately account for the different cross-section effects, the DOS has been obtained as 
a weighted sum of the $s$-orbital and $p$-orbital projected-DOS, with 
weights given by the relative cross-section of the $s$ and $p$ states at a phonon energy of 800~eV \cite{Yeh19851}.} 
The plasmonic polaron DOS has been magnified by a factor of two 
to enhance the visibility on the same scale of the quasiparticle spectral features. 
The experimental integrated x-ray photoemission spectrum (XPS) reproduced from Ref.~\onlinecite{PhysRevLett.107.166401} 
is reported for comparison in Fig.~\ref{fig:Sifull}~(b). 
The spectral function exhibits a set of bright quasiparticle bands 
for binding energies in the range of 0 to 12~eV. 
These bands arise from the first term of Eq.~(\ref{eq-cumulant-appr1}) and 
define the ordinary band structure of silicon. Their intensity is proportional to the emission rate 
of a photoelectron in absence of plasmonic excitations. 
In addition to the ordinary quasiparticle band structure, electron-plasmon coupling introduces 
additional dispersive spectral features in the angle-resolved spectrum. 
These spectral features, which manifest themselves as broadened replica of the valence 
band structure, reveal the excitation of plasmonic polarons, i.e., elementary excitation in which the energy 
of the absorbed photon contributes to the emission of a photoelectron and the excitation of a plasmon with 
energy $\sim\Omega_p$.

%
Our previous work\cite{Caruso2015} revealed that, similarly to ordinary quasiparticle bands, 
also plasmonic polaron band structures may lead to the formation of Van Hove singularities 
in the density of states (DOS).
For quasiparticle bands, the density of states $J$ may be expressed as: 
\begin{align}\label{eq-DOS}
J(\omega) = \frac{1}{4\pi^3}\sum_n\int dS_{\bf k} 
\frac{1}{|\nabla_{\bf k}\varepsilon_{n{\bf k}}|}
\end{align}
where $S_{\bf k}$ denotes the surface $\varepsilon_{n{\bf k}}={\rm constant}$.
Equation~(\ref{eq-DOS}) indicates that whenever the 
first momentum derivative of the quasiparticle bands $\varepsilon_{n{\bf k}}$ 
vanishes  (that is, $|\nabla_{\bf k}\varepsilon_{n{\bf k}}|=0$)
the density of states exhibits sharp resonances, known as Van Hove singularities.\cite{Kittel:ISSP} 
Consequently, the structures that characterize the DOS for binding energies between 0 and 
12~eV may be attributed to specific high-symmetry points of the first BZ, 
whereby quasiparticle bands are flat, and thus their derivative vanishes. 

  \begin{figure}
  \includegraphics[width=0.48\textwidth]{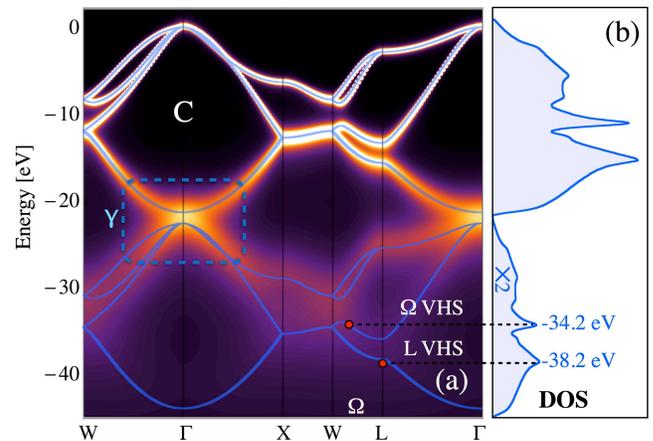}
  \caption{Spectral function (a) and DOS (b) of diamond determined 
  from the plasmonic polaron model [Eq.~(\ref{eq-cumulant-appr1})].}
  \label{fig:C}
  \end{figure}

The concept of Van Hove singularities may easily be generalized to the case of 
plasmonic polarons. In this case, however, we do not expect the emergence of sharp resonances 
owing to the large broadening characteristic of plasmonic polaron excitations. 
For silicon the XPS measurements at binding energies between 15 to 30~eV reveal 
a broad plasmon satellite characterized by a substructure of peaks and shoulders. 
These structure can be assigned to Van Hove singularities that arise 
from the plasmonic polaron band structures.

Having established a numerically efficient procedure to compute the plasmonic polaron contribution 
to the full spectral function, 
we can now proceed to assign the structures in the XPS plasmon satellite of silicon
to high-symmetry points in the first BZ.  
As illustrated in Fig.~\ref{fig:Sifull}, 
the two main peaks that characterize the plasmon satellite 
at 20 and 24~eV are Van Hove singularities resulting from a flattening 
of the plasmonic polaron bands at $X$ and $\Omega$ 
(the middle point on the $W$-$L$ high-symmetry line), respectively. 
Similarly, the two shoulders at 18 and 26~eV originate from the $L$ point.
The DOS of plasmonic polarons reported in Fig.~\ref{fig:Sifull}~(c) further validates this analysis,  
illustrating that the Van Hove singularities of the plasmonic polaron bands leads to spectral features in 
excellent agreement with the experimental reference data.
These results provide a simple rationale to unravel the complex features of 
the plasmon satellites in the integrated photoemission measurements of solids.
In particular, plasmon satellites 
can be attributed to plasmonic Van Hove singularities and their substructures 
arise from the flattening of the plasmonic polaron bands at 
different high-symmetry points in the Brillouin zone. 

%
We now move on to discuss the plasmonic polaron band structures of Ge and GaAs. 
In Fig.~\ref{fig:Ge/GaAs} we report the full spectral function of Ge 
(a) and GaAs (c) determined from the plasmonic polaron model and the corresponding DOS in panels (b) and (d), respectively. 
The spectral functions of Ge and GaAs exhibit similar qualitative features
to Si. For binding energies up to 15~eV below the Fermi energy, the spectrum is characterized 
by the ordinary quasiparticle band structure. 
For binding energies larger than 15~eV, the spectral function is dominated by the 
plasmonic polaron bands. These bands are expected to provide a dominant contribution 
to the density of states in this energy range.
Overall for Si, Ge, and GaAs, the broadening of the plasmonic polaron bands 
may obstruct the experimental observation of the individual bands. 
However, based on the results presented in Figs.~\ref{fig:Sifull} and \ref{fig:Ge/GaAs}, 
we identify general patterns that may drive the experimental observation 
of plasmonic polaron bands. In particular, Si, Ge, and GaAs manifest a depletion of density 
of states around the $\Gamma$ point for binding energies between 20 and 30~eV.
As illustrated in  Fig.~\ref{fig:Ge/GaAs}, this leads to 
the emergence of a diamond-shaped structure (denoted as $\gamma$ in Fig.~\ref{fig:Ge/GaAs}) 
in ARPES measurements. 
The low-intensity region $\gamma$ can be attributed to the 
gap between the first three degenerate plasmonic polaron bands (that is, the plasmonic 
replica of the three top-most quasiparticle bands) at the $\Gamma$ point
and the fourth plasmonic polaron band. 
Similarly, we observe a lowering of spectral intensity around $W$ and $X$ 
(denoted as $\alpha$ and $\beta$ in Fig.~\ref{fig:Ge/GaAs})
for binding energies between 11 and 18~eV below the Fermi energy. In this case, the $\alpha$ and $\beta$ regions 
stem from the gap between the lowest quasiparticle band and the first plasmonic polaron band.

Our calculations for the plasmonic polarons DOS [Fig.~\ref{fig:Ge/GaAs}~(b) and (d)] indicates that, similarly to the quasiparticle DOS, 
plasmonic polarons leads to the formation of well defined Van Hove singularities. 
We thus predict that XPS measurements of Ge (GaAs) should reveal a plasmon satellite 
characterized by at least three different structures at binding energies of 
18.5, 23.0, and 26.1~eV (18.1, 22.0, and 26.3~eV). As in the case of silicon, we 
attributed these structures to the flattening of the plasmonic polaron bands 
at different high-symmetry 
points of the Brillouin zone (the $X$, $\Omega$, and $L$ points)
as illustrated in panels (b) and (d) of Fig.~\ref{fig:Ge/GaAs}. 

As a last example, we consider the case of diamond.
The electronic properties of diamond are qualitatively different from those of the semiconductors 
discussed above. In particular, diamond is wide gap insulator ($E_{\rm g}\simeq5.5$~eV), 
its bands are more dispersive than Si, Ge, and GaAs, and it has a larger plasmon energy $\Omega_p\simeq22.5$~eV. 
These features are expected to affect considerably the band structures of plasmonic polarons.
An inspection of the full spectral function [Fig.~\ref{fig:C}~(a)] reveals that
the plasmonic polaron bands emerge at larger binding 
energies as compared to the other compounds, owing to the larger plasmon energy $\Omega_p$. 
Because of the shorter lifetime of plasmonic polarons at these binding energies, however, their spectral signatures 
appear broader and less intense. 
For the case of diamond, electron-plasmon interactions  are thus expected 
to introduce low intensity spectral features that might be 
difficult to reveal in ARPES measurements. 
However, a possible evidence of electron-plasmon coupling might arise in the form of an $X$-shaped resonance 
at the bottom of the valence band [denoted as $\gamma$ in Fig.~\ref{fig:C}~(a)] which results from the 
merging of the lowest-energy valence band and the highest-energy plasmonic polaron band at the $\Gamma$ point.
Similarly, the large bandwidth of quasiparticle and plasmonic polaron bands of 
diamonds leads to less pronounced Van Hove singularities as compared to silicon and GaAs [Fig.~\ref{fig:C}~(b)].
Correspondingly, the plasmon satellite is unlikely to exhibit substructures that may be 
attributed to Van Hove singularities occurring at different points in the Brillouin zone. 
In particular, the broadening is expected to wash out and merge the  
two resonances at 34.2 and 38.2~eV indicated in Fig.~\ref{fig:C}~(b). 

\section{Conclusions}
In summary, we presented a study of the spectroscopic signatures of electronic-plasmon coupling 
in integrated and angle-resolved photoemission experiments on tetrahedral semiconductors. 
Based on the cumulant expansion for the spectral function, 
we derived  the plasmonic polaron model, 
a simplified approach for the calculation of  
the band structures of plasmonic polarons in solids. 
This model, that does not require the explicit computation of the $GW$ self-energy, 
allows us to predict the full spectral function of solids at the numerical cost of a 
band structure calculation. 
Our results unveil the effects of electron-plasmon coupling for Si, Ge, GaAs, and diamond.
Si, Ge, and GaAs exhibit reasonably well defined plasmonic polaron band 
structures, that manifest themselves as 
broadened replica of the valence bands red-shifted by the plasmon energy, whereas for diamond plasmonic 
polarons spectral features are suppressed by short lifetime effects.
Additionally, this study reveals that the different structures that characterize 
the plasmon satellite of silicon may be attributed to {\it plasmonic Van Hove singularities} occurring at 
specific high-symmetry points in the Brillouin zone. This indicates that the dispersive nature of plasmonic polarons 
may be revealed already in integrated photoemission experiments.
Finally, our analysis provides clear guidelines for identifying the spectral signatures 
of plasmonic polarons in ARPES. 
In particular, for Si, Ge, and GaAs, plasmonic polaron bands should lead to low-intensity 
regions in the spectral function around the $X$ and $\Gamma$ points for binding energies 
around 15 and 25~eV, respectively. 

Overall, plasmonic polaron bands emerge as an important new concept for the interpretation 
of integrated and angle-resolved photoemission experiments. These findings calls for 
renewed experimental efforts along the lines of Ref.~\onlinecite{Lischner2015} for silicon.
The work presented here provides a simple predictive tool that may prove useful 
to unravel the complexity of photoemission spectra for a broad class of materials
(including, e.g., semiconductors, metals, and $d$/$f$-electron systems),
and to advance our understanding of elementary electronic excitations in condensed matter.

\acknowledgments
We thank H.~Lambert, B.~Gumhalter, J.~Lischner, and S.~G.~Louie for useful discussions.
This work was supported by the Leverhulme Trust (Grant RL-2012-001) and the European Research Council
(EU FP7 / ERC grant no. 239578 and EU FP7/grant no. 604391 Graphene Flagship). Calculations were
performed at the Oxford Supercomputing Centre and at the Oxford Materials Modelling Laboratory.


\begin{thebibliography}{45}%
\makeatletter
\providecommand \@ifxundefined [1]{%
 \@ifx{#1\undefined}
}%
\providecommand \@ifnum [1]{%
 \ifnum #1\expandafter \@firstoftwo
 \else \expandafter \@secondoftwo
 \fi
}%
\providecommand \@ifx [1]{%
 \ifx #1\expandafter \@firstoftwo
 \else \expandafter \@secondoftwo
 \fi
}%
\providecommand \natexlab [1]{#1}%
\providecommand \enquote  [1]{``#1''}%
\providecommand \bibnamefont  [1]{#1}%
\providecommand \bibfnamefont [1]{#1}%
\providecommand \citenamefont [1]{#1}%
\providecommand \href@noop [0]{\@secondoftwo}%
\providecommand \href [0]{\begingroup \@sanitize@url \@href}%
\providecommand \@href[1]{\@@startlink{#1}\@@href}%
\providecommand \@@href[1]{\endgroup#1\@@endlink}%
\providecommand \@sanitize@url [0]{\catcode `\\12\catcode `\$12\catcode
  `\&12\catcode `\#12\catcode `\^12\catcode `\_12\catcode `\%12\relax}%
\providecommand \@@startlink[1]{}%
\providecommand \@@endlink[0]{}%
\providecommand \url  [0]{\begingroup\@sanitize@url \@url }%
\providecommand \@url [1]{\endgroup\@href {#1}{\urlprefix }}%
\providecommand \urlprefix  [0]{URL }%
\providecommand \Eprint [0]{\href }%
\providecommand \doibase [0]{http://dx.doi.org/}%
\providecommand \selectlanguage [0]{\@gobble}%
\providecommand \bibinfo  [0]{\@secondoftwo}%
\providecommand \bibfield  [0]{\@secondoftwo}%
\providecommand \translation [1]{[#1]}%
\providecommand \BibitemOpen [0]{}%
\providecommand \bibitemStop [0]{}%
\providecommand \bibitemNoStop [0]{.\EOS\space}%
\providecommand \EOS [0]{\spacefactor3000\relax}%
\providecommand \BibitemShut  [1]{\csname bibitem#1\endcsname}%
\let\auto@bib@innerbib\@empty
\bibitem [{\citenamefont {Cooper}(1956)}]{Cooper1956}%
  \BibitemOpen
  \bibfield  {author} {\bibinfo {author} {\bibfnamefont {L.~N.}\ \bibnamefont
  {Cooper}},\ }\href {\doibase 10.1103/PhysRev.104.1189} {\bibfield  {journal}
  {\bibinfo  {journal} {Phys. Rev.}\ }\textbf {\bibinfo {volume} {104}},\
  \bibinfo {pages} {1189} (\bibinfo {year} {1956})}\BibitemShut {NoStop}%
\bibitem [{\citenamefont {Bardeen}\ \emph {et~al.}(1957)\citenamefont
  {Bardeen}, \citenamefont {Cooper},\ and\ \citenamefont
  {Schrieffer}}]{BCS1957}%
  \BibitemOpen
  \bibfield  {author} {\bibinfo {author} {\bibfnamefont {J.}~\bibnamefont
  {Bardeen}}, \bibinfo {author} {\bibfnamefont {L.~N.}\ \bibnamefont {Cooper}},
  \ and\ \bibinfo {author} {\bibfnamefont {J.~R.}\ \bibnamefont {Schrieffer}},\
  }\href {\doibase 10.1103/PhysRev.106.162} {\bibfield  {journal} {\bibinfo
  {journal} {Phys. Rev.}\ }\textbf {\bibinfo {volume} {106}},\ \bibinfo {pages}
  {162} (\bibinfo {year} {1957})}\BibitemShut {NoStop}%
\bibitem [{\citenamefont {Hedin}(1965)}]{PhysRev.139.A796}%
  \BibitemOpen
  \bibfield  {author} {\bibinfo {author} {\bibfnamefont {L.}~\bibnamefont
  {Hedin}},\ }\href {\doibase 10.1103/PhysRev.139.A796} {\bibfield  {journal}
  {\bibinfo  {journal} {Phys. Rev.}\ }\textbf {\bibinfo {volume} {139}},\
  \bibinfo {pages} {A796} (\bibinfo {year} {1965})}\BibitemShut {NoStop}%
\bibitem [{\citenamefont {Hybertsen}\ and\ \citenamefont
  {Louie}(1986)}]{Hybertsen1986}%
  \BibitemOpen
  \bibfield  {author} {\bibinfo {author} {\bibfnamefont {M.~S.}\ \bibnamefont
  {Hybertsen}}\ and\ \bibinfo {author} {\bibfnamefont {S.~G.}\ \bibnamefont
  {Louie}},\ }\href {\doibase 10.1103/PhysRevB.34.5390} {\bibfield  {journal}
  {\bibinfo  {journal} {Phys. Rev. B}\ }\textbf {\bibinfo {volume} {34}},\
  \bibinfo {pages} {5390} (\bibinfo {year} {1986})}\BibitemShut {NoStop}%
\bibitem [{\citenamefont {Lundqvist}(1967{\natexlab{a}})}]{Lundqvist}%
  \BibitemOpen
  \bibfield  {author} {\bibinfo {author} {\bibfnamefont {B.}~\bibnamefont
  {Lundqvist}},\ }\href {\doibase 10.1007/BF02422716} {\bibfield  {journal}
  {\bibinfo  {journal} {Phys. Kondens. Mater.}\ }\textbf {\bibinfo {volume}
  {6}},\ \bibinfo {pages} {193} (\bibinfo {year}
  {1967}{\natexlab{a}})}\BibitemShut {NoStop}%
\bibitem [{\citenamefont {Langreth}(1970)}]{PhysRevB.1.471}%
  \BibitemOpen
  \bibfield  {author} {\bibinfo {author} {\bibfnamefont {D.~C.}\ \bibnamefont
  {Langreth}},\ }\href {\doibase 10.1103/PhysRevB.1.471} {\bibfield  {journal}
  {\bibinfo  {journal} {Phys. Rev. B}\ }\textbf {\bibinfo {volume} {1}},\
  \bibinfo {pages} {471} (\bibinfo {year} {1970})}\BibitemShut {NoStop}%
\bibitem [{\citenamefont {Gumhalter}(2012)}]{Gumhalter2012163}%
  \BibitemOpen
  \bibfield  {author} {\bibinfo {author} {\bibfnamefont {B.}~\bibnamefont
  {Gumhalter}},\ }\href {\doibase
  http://dx.doi.org/10.1016/j.progsurf.2012.05.004} {\bibfield  {journal}
  {\bibinfo  {journal} {Progress in Surface Science}\ }\textbf {\bibinfo
  {volume} {87}},\ \bibinfo {pages} {163 } (\bibinfo {year}
  {2012})}\BibitemShut {NoStop}%
\bibitem [{\citenamefont {Gumhalter}(2005)}]{PhysRevB.72.165406}%
  \BibitemOpen
  \bibfield  {author} {\bibinfo {author} {\bibfnamefont {B.}~\bibnamefont
  {Gumhalter}},\ }\href {\doibase 10.1103/PhysRevB.72.165406} {\bibfield
  {journal} {\bibinfo  {journal} {Phys. Rev. B}\ }\textbf {\bibinfo {volume}
  {72}},\ \bibinfo {pages} {165406} (\bibinfo {year} {2005})}\BibitemShut
  {NoStop}%
\bibitem [{\citenamefont {Holm}\ and\ \citenamefont
  {Aryasetiawan}(1997)}]{Holm1997}%
  \BibitemOpen
  \bibfield  {author} {\bibinfo {author} {\bibfnamefont {B.}~\bibnamefont
  {Holm}}\ and\ \bibinfo {author} {\bibfnamefont {F.}~\bibnamefont
  {Aryasetiawan}},\ }\href {\doibase 10.1103/PhysRevB.56.12825} {\bibfield
  {journal} {\bibinfo  {journal} {Phys. Rev. B}\ }\textbf {\bibinfo {volume}
  {56}},\ \bibinfo {pages} {12825} (\bibinfo {year} {1997})}\BibitemShut
  {NoStop}%
\bibitem [{\citenamefont {Kas}\ \emph {et~al.}(2014)\citenamefont {Kas},
  \citenamefont {Rehr},\ and\ \citenamefont {Reining}}]{Kas2014}%
  \BibitemOpen
  \bibfield  {author} {\bibinfo {author} {\bibfnamefont {J.~J.}\ \bibnamefont
  {Kas}}, \bibinfo {author} {\bibfnamefont {J.~J.}\ \bibnamefont {Rehr}}, \
  and\ \bibinfo {author} {\bibfnamefont {L.}~\bibnamefont {Reining}},\ }\href
  {\doibase 10.1103/PhysRevB.90.085112} {\bibfield  {journal} {\bibinfo
  {journal} {Phys. Rev. B}\ }\textbf {\bibinfo {volume} {90}},\ \bibinfo
  {pages} {085112} (\bibinfo {year} {2014})}\BibitemShut {NoStop}%
\bibitem [{\citenamefont {Aryasetiawan}\ \emph {et~al.}(1996)\citenamefont
  {Aryasetiawan}, \citenamefont {Hedin},\ and\ \citenamefont
  {Karlsson}}]{PhysRevLett.77.2268}%
  \BibitemOpen
  \bibfield  {author} {\bibinfo {author} {\bibfnamefont {F.}~\bibnamefont
  {Aryasetiawan}}, \bibinfo {author} {\bibfnamefont {L.}~\bibnamefont {Hedin}},
  \ and\ \bibinfo {author} {\bibfnamefont {K.}~\bibnamefont {Karlsson}},\
  }\href {\doibase 10.1103/PhysRevLett.77.2268} {\bibfield  {journal} {\bibinfo
   {journal} {Phys. Rev. Lett.}\ }\textbf {\bibinfo {volume} {77}},\ \bibinfo
  {pages} {2268} (\bibinfo {year} {1996})}\BibitemShut {NoStop}%
\bibitem [{\citenamefont {Guzzo}\ \emph {et~al.}(2012)\citenamefont {Guzzo}
  \emph {et~al.}}]{Guzzo2012}%
  \BibitemOpen
  \bibfield  {author} {\bibinfo {author} {\bibfnamefont {M.}~\bibnamefont
  {Guzzo}} \emph {et~al.},\ }\href
  {http://dx.doi.org/10.1140/epjb/e2012-30267-y} {\bibfield  {journal}
  {\bibinfo  {journal} {Eur. Phys. J. B}\ }\textbf {\bibinfo {volume} {85}},\
  \bibinfo {eid} {324} (\bibinfo {year} {2012})}\BibitemShut {NoStop}%
\bibitem [{\citenamefont {Guzzo}\ \emph {et~al.}(2014)\citenamefont {Guzzo}
  \emph {et~al.}}]{Guzzo2014}%
  \BibitemOpen
  \bibfield  {author} {\bibinfo {author} {\bibfnamefont {M.}~\bibnamefont
  {Guzzo}} \emph {et~al.},\ }\href {\doibase 10.1103/PhysRevB.89.085425}
  {\bibfield  {journal} {\bibinfo  {journal} {Phys. Rev. B}\ }\textbf {\bibinfo
  {volume} {89}},\ \bibinfo {pages} {085425} (\bibinfo {year}
  {2014})}\BibitemShut {NoStop}%
\bibitem [{\citenamefont {Lischner}\ \emph {et~al.}(2013)\citenamefont
  {Lischner}, \citenamefont {Vigil-Fowler},\ and\ \citenamefont
  {Louie}}]{PhysRevLett.110.146801}%
  \BibitemOpen
  \bibfield  {author} {\bibinfo {author} {\bibfnamefont {J.}~\bibnamefont
  {Lischner}}, \bibinfo {author} {\bibfnamefont {D.}~\bibnamefont
  {Vigil-Fowler}}, \ and\ \bibinfo {author} {\bibfnamefont {S.~G.}\
  \bibnamefont {Louie}},\ }\href {\doibase 10.1103/PhysRevLett.110.146801}
  {\bibfield  {journal} {\bibinfo  {journal} {Phys. Rev. Lett.}\ }\textbf
  {\bibinfo {volume} {110}},\ \bibinfo {pages} {146801} (\bibinfo {year}
  {2013})}\BibitemShut {NoStop}%
\bibitem [{\citenamefont {Caruso}\ \emph {et~al.}(2015)\citenamefont {Caruso},
  \citenamefont {Lambert},\ and\ \citenamefont {Giustino}}]{Caruso2015}%
  \BibitemOpen
  \bibfield  {author} {\bibinfo {author} {\bibfnamefont {F.}~\bibnamefont
  {Caruso}}, \bibinfo {author} {\bibfnamefont {H.}~\bibnamefont {Lambert}}, \
  and\ \bibinfo {author} {\bibfnamefont {F.}~\bibnamefont {Giustino}},\ }\href
  {\doibase 10.1103/PhysRevLett.114.146404} {\bibfield  {journal} {\bibinfo
  {journal} {Phys. Rev. Lett.}\ }\textbf {\bibinfo {volume} {114}},\ \bibinfo
  {pages} {146404} (\bibinfo {year} {2015})}\BibitemShut {NoStop}%
\bibitem [{\citenamefont {Lischner}\ \emph {et~al.}(2015)\citenamefont
  {Lischner}, \citenamefont {P\'alsson}, \citenamefont {Vigil-Fowler},
  \citenamefont {Nemsak}, \citenamefont {Avila}, \citenamefont {Asensio},
  \citenamefont {Fadley},\ and\ \citenamefont {Louie}}]{Lischner2015}%
  \BibitemOpen
  \bibfield  {author} {\bibinfo {author} {\bibfnamefont {J.}~\bibnamefont
  {Lischner}}, \bibinfo {author} {\bibfnamefont {G.~K.}\ \bibnamefont
  {P\'alsson}}, \bibinfo {author} {\bibfnamefont {D.}~\bibnamefont
  {Vigil-Fowler}}, \bibinfo {author} {\bibfnamefont {S.}~\bibnamefont
  {Nemsak}}, \bibinfo {author} {\bibfnamefont {J.}~\bibnamefont {Avila}},
  \bibinfo {author} {\bibfnamefont {M.~C.}\ \bibnamefont {Asensio}}, \bibinfo
  {author} {\bibfnamefont {C.~S.}\ \bibnamefont {Fadley}}, \ and\ \bibinfo
  {author} {\bibfnamefont {S.~G.}\ \bibnamefont {Louie}},\ }\href {\doibase
  10.1103/PhysRevB.91.205113} {\bibfield  {journal} {\bibinfo  {journal} {Phys.
  Rev. B}\ }\textbf {\bibinfo {volume} {91}},\ \bibinfo {pages} {205113}
  (\bibinfo {year} {2015})}\BibitemShut {NoStop}%
\bibitem [{\citenamefont {Hedin}(1980)}]{1402-4896-21-3-4-039}%
  \BibitemOpen
  \bibfield  {author} {\bibinfo {author} {\bibfnamefont {L.}~\bibnamefont
  {Hedin}},\ }\href {http://iopscience.iop.org/1402-4896/21/3-4/039} {\bibfield
   {journal} {\bibinfo  {journal} {Phys. Scr.}\ }\textbf {\bibinfo {volume}
  {21}},\ \bibinfo {pages} {477} (\bibinfo {year} {1980})}\BibitemShut
  {NoStop}%
\bibitem [{\citenamefont {Guzzo}\ \emph {et~al.}(2011)\citenamefont {Guzzo},
  \citenamefont {Lani}, \citenamefont {Sottile}, \citenamefont {Romaniello},
  \citenamefont {Gatti}, \citenamefont {Kas}, \citenamefont {Rehr},
  \citenamefont {Silly}, \citenamefont {Sirotti},\ and\ \citenamefont
  {Reining}}]{PhysRevLett.107.166401}%
  \BibitemOpen
  \bibfield  {author} {\bibinfo {author} {\bibfnamefont {M.}~\bibnamefont
  {Guzzo}}, \bibinfo {author} {\bibfnamefont {G.}~\bibnamefont {Lani}},
  \bibinfo {author} {\bibfnamefont {F.}~\bibnamefont {Sottile}}, \bibinfo
  {author} {\bibfnamefont {P.}~\bibnamefont {Romaniello}}, \bibinfo {author}
  {\bibfnamefont {M.}~\bibnamefont {Gatti}}, \bibinfo {author} {\bibfnamefont
  {J.~J.}\ \bibnamefont {Kas}}, \bibinfo {author} {\bibfnamefont {J.~J.}\
  \bibnamefont {Rehr}}, \bibinfo {author} {\bibfnamefont {M.~G.}\ \bibnamefont
  {Silly}}, \bibinfo {author} {\bibfnamefont {F.}~\bibnamefont {Sirotti}}, \
  and\ \bibinfo {author} {\bibfnamefont {L.}~\bibnamefont {Reining}},\ }\href
  {\doibase 10.1103/PhysRevLett.107.166401} {\bibfield  {journal} {\bibinfo
  {journal} {Phys. Rev. Lett.}\ }\textbf {\bibinfo {volume} {107}},\ \bibinfo
  {pages} {166401} (\bibinfo {year} {2011})}\BibitemShut {NoStop}%
\bibitem [{\citenamefont {Hohenberg}\ and\ \citenamefont
  {Kohn}(1964)}]{PhysRev.136.B864}%
  \BibitemOpen
  \bibfield  {author} {\bibinfo {author} {\bibfnamefont {P.}~\bibnamefont
  {Hohenberg}}\ and\ \bibinfo {author} {\bibfnamefont {W.}~\bibnamefont
  {Kohn}},\ }\href {\doibase 10.1103/PhysRev.136.B864} {\bibfield  {journal}
  {\bibinfo  {journal} {Phys. Rev.}\ }\textbf {\bibinfo {volume} {136}},\
  \bibinfo {pages} {B864} (\bibinfo {year} {1964})}\BibitemShut {NoStop}%
\bibitem [{\citenamefont {Kohn}\ and\ \citenamefont
  {Sham}(1965)}]{PhysRev.140.A1133}%
  \BibitemOpen
  \bibfield  {author} {\bibinfo {author} {\bibfnamefont {W.}~\bibnamefont
  {Kohn}}\ and\ \bibinfo {author} {\bibfnamefont {L.~J.}\ \bibnamefont
  {Sham}},\ }\href {\doibase 10.1103/PhysRev.140.A1133} {\bibfield  {journal}
  {\bibinfo  {journal} {Phys. Rev.}\ }\textbf {\bibinfo {volume} {140}},\
  \bibinfo {pages} {A1133} (\bibinfo {year} {1965})}\BibitemShut {NoStop}%
\bibitem [{\citenamefont {Giustino}\ \emph {et~al.}(2010)\citenamefont
  {Giustino}, \citenamefont {Cohen},\ and\ \citenamefont
  {Louie}}]{PhysRevB.81.115105}%
  \BibitemOpen
  \bibfield  {author} {\bibinfo {author} {\bibfnamefont {F.}~\bibnamefont
  {Giustino}}, \bibinfo {author} {\bibfnamefont {M.~L.}\ \bibnamefont {Cohen}},
  \ and\ \bibinfo {author} {\bibfnamefont {S.~G.}\ \bibnamefont {Louie}},\
  }\href {\doibase 10.1103/PhysRevB.81.115105} {\bibfield  {journal} {\bibinfo
  {journal} {Phys. Rev. B}\ }\textbf {\bibinfo {volume} {81}},\ \bibinfo
  {pages} {115105} (\bibinfo {year} {2010})}\BibitemShut {NoStop}%
\bibitem [{\citenamefont {Lundqvist}(1967{\natexlab{b}})}]{Lundqvist2}%
  \BibitemOpen
  \bibfield  {author} {\bibinfo {author} {\bibfnamefont {B.}~\bibnamefont
  {Lundqvist}},\ }\href {\doibase 10.1007/BF02422717} {\bibfield  {journal}
  {\bibinfo  {journal} {Physik der kondensierten Materie}\ }\textbf {\bibinfo
  {volume} {6}},\ \bibinfo {pages} {206} (\bibinfo {year}
  {1967}{\natexlab{b}})}\BibitemShut {NoStop}%
\bibitem [{\citenamefont {Godby}\ and\ \citenamefont
  {Needs}(1989)}]{Godby1989}%
  \BibitemOpen
  \bibfield  {author} {\bibinfo {author} {\bibfnamefont {R.~W.}\ \bibnamefont
  {Godby}}\ and\ \bibinfo {author} {\bibfnamefont {R.~J.}\ \bibnamefont
  {Needs}},\ }\href {\doibase 10.1103/PhysRevLett.62.1169} {\bibfield
  {journal} {\bibinfo  {journal} {Phys. Rev. Lett.}\ }\textbf {\bibinfo
  {volume} {62}},\ \bibinfo {pages} {1169} (\bibinfo {year}
  {1989})}\BibitemShut {NoStop}%
\bibitem [{\citenamefont {Engel}\ and\ \citenamefont
  {Farid}(1993)}]{Engel1993}%
  \BibitemOpen
  \bibfield  {author} {\bibinfo {author} {\bibfnamefont {G.~E.}\ \bibnamefont
  {Engel}}\ and\ \bibinfo {author} {\bibfnamefont {B.}~\bibnamefont {Farid}},\
  }\href {\doibase 10.1103/PhysRevB.47.15931} {\bibfield  {journal} {\bibinfo
  {journal} {Phys. Rev. B}\ }\textbf {\bibinfo {volume} {47}},\ \bibinfo
  {pages} {15931} (\bibinfo {year} {1993})}\BibitemShut {NoStop}%
\bibitem [{\citenamefont {Kas}\ \emph {et~al.}(2007)\citenamefont {Kas},
  \citenamefont {Sorini}, \citenamefont {Prange}, \citenamefont {Cambell},
  \citenamefont {Soininen},\ and\ \citenamefont {Rehr}}]{PhysRevB.76.195116}%
  \BibitemOpen
  \bibfield  {author} {\bibinfo {author} {\bibfnamefont {J.~J.}\ \bibnamefont
  {Kas}}, \bibinfo {author} {\bibfnamefont {A.~P.}\ \bibnamefont {Sorini}},
  \bibinfo {author} {\bibfnamefont {M.~P.}\ \bibnamefont {Prange}}, \bibinfo
  {author} {\bibfnamefont {L.~W.}\ \bibnamefont {Cambell}}, \bibinfo {author}
  {\bibfnamefont {J.~A.}\ \bibnamefont {Soininen}}, \ and\ \bibinfo {author}
  {\bibfnamefont {J.~J.}\ \bibnamefont {Rehr}},\ }\href {\doibase
  10.1103/PhysRevB.76.195116} {\bibfield  {journal} {\bibinfo  {journal} {Phys.
  Rev. B}\ }\textbf {\bibinfo {volume} {76}},\ \bibinfo {pages} {195116}
  (\bibinfo {year} {2007})}\BibitemShut {NoStop}%
\bibitem [{\citenamefont {Stankovski}\ \emph {et~al.}(2011)\citenamefont
  {Stankovski}, \citenamefont {Antonius}, \citenamefont {Waroquiers},
  \citenamefont {Miglio}, \citenamefont {Dixit}, \citenamefont {Sankaran},
  \citenamefont {Giantomassi}, \citenamefont {Gonze}, \citenamefont
  {C\^ot\'e},\ and\ \citenamefont {Rignanese}}]{PhysRevB.84.241201}%
  \BibitemOpen
  \bibfield  {author} {\bibinfo {author} {\bibfnamefont {M.}~\bibnamefont
  {Stankovski}}, \bibinfo {author} {\bibfnamefont {G.}~\bibnamefont
  {Antonius}}, \bibinfo {author} {\bibfnamefont {D.}~\bibnamefont
  {Waroquiers}}, \bibinfo {author} {\bibfnamefont {A.}~\bibnamefont {Miglio}},
  \bibinfo {author} {\bibfnamefont {H.}~\bibnamefont {Dixit}}, \bibinfo
  {author} {\bibfnamefont {K.}~\bibnamefont {Sankaran}}, \bibinfo {author}
  {\bibfnamefont {M.}~\bibnamefont {Giantomassi}}, \bibinfo {author}
  {\bibfnamefont {X.}~\bibnamefont {Gonze}}, \bibinfo {author} {\bibfnamefont
  {M.}~\bibnamefont {C\^ot\'e}}, \ and\ \bibinfo {author} {\bibfnamefont
  {G.-M.}\ \bibnamefont {Rignanese}},\ }\href {\doibase
  10.1103/PhysRevB.84.241201} {\bibfield  {journal} {\bibinfo  {journal} {Phys.
  Rev. B}\ }\textbf {\bibinfo {volume} {84}},\ \bibinfo {pages} {241201}
  (\bibinfo {year} {2011})}\BibitemShut {NoStop}%
\bibitem [{\citenamefont {Larson}\ \emph {et~al.}(2013)\citenamefont {Larson},
  \citenamefont {Dvorak},\ and\ \citenamefont {Wu}}]{PhysRevB.88.125205}%
  \BibitemOpen
  \bibfield  {author} {\bibinfo {author} {\bibfnamefont {P.}~\bibnamefont
  {Larson}}, \bibinfo {author} {\bibfnamefont {M.}~\bibnamefont {Dvorak}}, \
  and\ \bibinfo {author} {\bibfnamefont {Z.}~\bibnamefont {Wu}},\ }\href
  {\doibase 10.1103/PhysRevB.88.125205} {\bibfield  {journal} {\bibinfo
  {journal} {Phys. Rev. B}\ }\textbf {\bibinfo {volume} {88}},\ \bibinfo
  {pages} {125205} (\bibinfo {year} {2013})}\BibitemShut {NoStop}%
\bibitem [{\citenamefont {Shih}\ \emph {et~al.}(2010)\citenamefont {Shih},
  \citenamefont {Xue}, \citenamefont {Zhang}, \citenamefont {Cohen},\ and\
  \citenamefont {Louie}}]{PhysRevLett.105.146401}%
  \BibitemOpen
  \bibfield  {author} {\bibinfo {author} {\bibfnamefont {B.-C.}\ \bibnamefont
  {Shih}}, \bibinfo {author} {\bibfnamefont {Y.}~\bibnamefont {Xue}}, \bibinfo
  {author} {\bibfnamefont {P.}~\bibnamefont {Zhang}}, \bibinfo {author}
  {\bibfnamefont {M.~L.}\ \bibnamefont {Cohen}}, \ and\ \bibinfo {author}
  {\bibfnamefont {S.~G.}\ \bibnamefont {Louie}},\ }\href {\doibase
  10.1103/PhysRevLett.105.146401} {\bibfield  {journal} {\bibinfo  {journal}
  {Phys. Rev. Lett.}\ }\textbf {\bibinfo {volume} {105}},\ \bibinfo {pages}
  {146401} (\bibinfo {year} {2010})}\BibitemShut {NoStop}%
\bibitem [{Note1()}]{Note1}%
  \BibitemOpen
  \bibinfo {note} {The same expression can be derived combining the Lorentzian
  approximation for the self-energy with the expression $Z_{n{\protect \bf
  k}}=\left [1+\DOTSI \intop \ilimits@ \protect \frac {\beta _{n{\protect \bf
  k}}(\omega )}{\omega ^2}d\omega \right ]^{-1}$ derived in Ref.~\protect
  \rev@citealpnum {0953-8984-11-42-201}.}\BibitemShut {Stop}%
\bibitem [{\citenamefont {Tosatti}\ and\ \citenamefont
  {Parravicini}(1971)}]{Tosatti1971623}%
  \BibitemOpen
  \bibfield  {author} {\bibinfo {author} {\bibfnamefont {E.}~\bibnamefont
  {Tosatti}}\ and\ \bibinfo {author} {\bibfnamefont {G.~P.}\ \bibnamefont
  {Parravicini}},\ }\href {\doibase
  http://dx.doi.org/10.1016/0022-3697(71)90011-4} {\bibfield  {journal}
  {\bibinfo  {journal} {Journal of Physics and Chemistry of Solids}\ }\textbf
  {\bibinfo {volume} {32}},\ \bibinfo {pages} {623 } (\bibinfo {year}
  {1971})}\BibitemShut {NoStop}%
\bibitem [{\citenamefont {Giannozzi}\ \emph {et~al.}(2009)\citenamefont
  {Giannozzi} \emph {et~al.}}]{0953-8984-21-39-395502}%
  \BibitemOpen
  \bibfield  {author} {\bibinfo {author} {\bibfnamefont {P.}~\bibnamefont
  {Giannozzi}} \emph {et~al.},\ }\href
  {http://stacks.iop.org/0953-8984/21/i=39/a=395502} {\bibfield  {journal}
  {\bibinfo  {journal} {J. Phys.: Condens. Matter}\ }\textbf {\bibinfo {volume}
  {21}},\ \bibinfo {pages} {395502} (\bibinfo {year} {2009})}\BibitemShut
  {NoStop}%
\bibitem [{Note2()}]{Note2}%
  \BibitemOpen
  \bibinfo {note} {We employed the Perdew-Burke-Ernzerhof (PBE) version of the
  generalized-gradient approximation\cite {PhysRevLett.77.3865} for the
  Kohn-Sham DFT exchange-correlation functional.\cite
  {PhysRev.136.B864,PhysRev.140.A1133} Our atomistic models of the Si, Ge,
  GaAs, and diamond crystal structures are based on the experimental structural
  parameters. Only valence electrons are treated explicitly in our
  calculations, whereas core electrons are accounted for through
  Rappe-Rabe-Kaxiras-Joannopoulos ultrasoft pseudopotentials.\cite
  {PhysRevB.41.1227} The momentum integrals over the first Brillouin zone are
  discretized over a $8\times 8\times 8$ homogeneous Monkhorst-Pack grid, and
  all plane waves up to a kinetic energy cutoff of 30~Ry are included in the
  calculation.}\BibitemShut {Stop}%
\bibitem [{\citenamefont {Nozi\`eres}\ and\ \citenamefont
  {Pines}(1959)}]{PhysRev.113.1254}%
  \BibitemOpen
  \bibfield  {author} {\bibinfo {author} {\bibfnamefont {P.}~\bibnamefont
  {Nozi\`eres}}\ and\ \bibinfo {author} {\bibfnamefont {D.}~\bibnamefont
  {Pines}},\ }\href {\doibase 10.1103/PhysRev.113.1254} {\bibfield  {journal}
  {\bibinfo  {journal} {Phys. Rev.}\ }\textbf {\bibinfo {volume} {113}},\
  \bibinfo {pages} {1254} (\bibinfo {year} {1959})}\BibitemShut {NoStop}%
\bibitem [{\citenamefont {Bechstedt}\ \emph {et~al.}(1997)\citenamefont
  {Bechstedt}, \citenamefont {Tenelsen}, \citenamefont {Adolph},\ and\
  \citenamefont {Del~Sole}}]{PhysRevLett.78.1528}%
  \BibitemOpen
  \bibfield  {author} {\bibinfo {author} {\bibfnamefont {F.}~\bibnamefont
  {Bechstedt}}, \bibinfo {author} {\bibfnamefont {K.}~\bibnamefont {Tenelsen}},
  \bibinfo {author} {\bibfnamefont {B.}~\bibnamefont {Adolph}}, \ and\ \bibinfo
  {author} {\bibfnamefont {R.}~\bibnamefont {Del~Sole}},\ }\href {\doibase
  10.1103/PhysRevLett.78.1528} {\bibfield  {journal} {\bibinfo  {journal}
  {Phys. Rev. Lett.}\ }\textbf {\bibinfo {volume} {78}},\ \bibinfo {pages}
  {1528} (\bibinfo {year} {1997})}\BibitemShut {NoStop}%
\bibitem [{\citenamefont {Hedin}\ \emph {et~al.}(1998)\citenamefont {Hedin},
  \citenamefont {Michiels},\ and\ \citenamefont {Inglesfield}}]{Hedin1998}%
  \BibitemOpen
  \bibfield  {author} {\bibinfo {author} {\bibfnamefont {L.}~\bibnamefont
  {Hedin}}, \bibinfo {author} {\bibfnamefont {J.}~\bibnamefont {Michiels}}, \
  and\ \bibinfo {author} {\bibfnamefont {J.}~\bibnamefont {Inglesfield}},\
  }\href {\doibase 10.1103/PhysRevB.58.15565} {\bibfield  {journal} {\bibinfo
  {journal} {Phys. Rev. B}\ }\textbf {\bibinfo {volume} {58}},\ \bibinfo
  {pages} {15565} (\bibinfo {year} {1998})}\BibitemShut {NoStop}%
\bibitem [{\citenamefont {Rehr}\ and\ \citenamefont {Albers}(2000)}]{Rehr2000}%
  \BibitemOpen
  \bibfield  {author} {\bibinfo {author} {\bibfnamefont {J.~J.}\ \bibnamefont
  {Rehr}}\ and\ \bibinfo {author} {\bibfnamefont {R.~C.}\ \bibnamefont
  {Albers}},\ }\href {\doibase 10.1103/RevModPhys.72.621} {\bibfield  {journal}
  {\bibinfo  {journal} {Rev. Mod. Phys.}\ }\textbf {\bibinfo {volume} {72}},\
  \bibinfo {pages} {621} (\bibinfo {year} {2000})}\BibitemShut {NoStop}%
\bibitem [{\citenamefont {Pines}(1999)}]{pines1999elementary}%
  \BibitemOpen
  \bibfield  {author} {\bibinfo {author} {\bibfnamefont {D.}~\bibnamefont
  {Pines}},\ }\href {https://books.google.co.uk/books?id=2Vsh\_KqV\_-4C} {\emph
  {\bibinfo {title} {Elementary Excitations in Solids: Lectures on Protons,
  Electrons, and Plasmons}}},\ Advanced book classics\ (\bibinfo  {publisher}
  {Advanced Book Program, Perseus Books},\ \bibinfo {year} {1999})\BibitemShut
  {NoStop}%
\bibitem [{\citenamefont {Mahan}(2000)}]{mahan2000many}%
  \BibitemOpen
  \bibfield  {author} {\bibinfo {author} {\bibfnamefont {G.}~\bibnamefont
  {Mahan}},\ }\href {http://books.google.co.uk/books?id=xzSgZ4-yyMEC} {\emph
  {\bibinfo {title} {Many-Particle Physics}}}\ (\bibinfo  {publisher}
  {Springer},\ \bibinfo {year} {2000})\BibitemShut {NoStop}%
\bibitem [{\citenamefont {Lambert}\ and\ \citenamefont
  {Giustino}(2013)}]{PhysRevB.88.075117}%
  \BibitemOpen
  \bibfield  {author} {\bibinfo {author} {\bibfnamefont {H.}~\bibnamefont
  {Lambert}}\ and\ \bibinfo {author} {\bibfnamefont {F.}~\bibnamefont
  {Giustino}},\ }\href {\doibase 10.1103/PhysRevB.88.075117} {\bibfield
  {journal} {\bibinfo  {journal} {Phys. Rev. B}\ }\textbf {\bibinfo {volume}
  {88}},\ \bibinfo {pages} {075117} (\bibinfo {year} {2013})}\BibitemShut
  {NoStop}%
\bibitem [{\citenamefont {Baroni}\ \emph {et~al.}(2001)\citenamefont {Baroni}
  \emph {et~al.}}]{baroni01}%
  \BibitemOpen
  \bibfield  {author} {\bibinfo {author} {\bibfnamefont {S.}~\bibnamefont
  {Baroni}} \emph {et~al.},\ }\href@noop {} {\bibfield  {journal} {\bibinfo
  {journal} {Rev. Mod. Phys.}\ }\textbf {\bibinfo {volume} {73}},\ \bibinfo
  {pages} {515} (\bibinfo {year} {2001})}\BibitemShut {NoStop}%
\bibitem [{\citenamefont {Yeh}\ and\ \citenamefont {Lindau}(1985)}]{Yeh19851}%
  \BibitemOpen
  \bibfield  {author} {\bibinfo {author} {\bibfnamefont {J.}~\bibnamefont
  {Yeh}}\ and\ \bibinfo {author} {\bibfnamefont {I.}~\bibnamefont {Lindau}},\
  }\href {\doibase http://dx.doi.org/10.1016/0092-640X(85)90016-6} {\bibfield
  {journal} {\bibinfo  {journal} {Atomic Data and Nuclear Data Tables}\
  }\textbf {\bibinfo {volume} {32}},\ \bibinfo {pages} {1 } (\bibinfo {year}
  {1985})}\BibitemShut {NoStop}%
\bibitem [{\citenamefont {Kittel}(1986)}]{Kittel:ISSP}%
  \BibitemOpen
  \bibfield  {author} {\bibinfo {author} {\bibfnamefont {C.}~\bibnamefont
  {Kittel}},\ }\href@noop {} {\emph {\bibinfo {title} {{Introduction to Solid
  State Physics}}}},\ \bibinfo {edition} {6th}\ ed.\ (\bibinfo  {publisher}
  {John Wiley \& Sons, Inc.},\ \bibinfo {address} {New York},\ \bibinfo {year}
  {1986})\BibitemShut {NoStop}%
\bibitem [{\citenamefont {Hedin}(1999)}]{0953-8984-11-42-201}%
  \BibitemOpen
  \bibfield  {author} {\bibinfo {author} {\bibfnamefont {L.}~\bibnamefont
  {Hedin}},\ }\href {http://stacks.iop.org/0953-8984/11/i=42/a=201} {\bibfield
  {journal} {\bibinfo  {journal} {J. Phys.: Condens. Matter}\ }\textbf
  {\bibinfo {volume} {11}},\ \bibinfo {pages} {R489} (\bibinfo {year}
  {1999})}\BibitemShut {NoStop}%
\bibitem [{\citenamefont {Perdew}\ \emph {et~al.}(1996)\citenamefont {Perdew},
  \citenamefont {Burke},\ and\ \citenamefont
  {Ernzerhof}}]{PhysRevLett.77.3865}%
  \BibitemOpen
  \bibfield  {author} {\bibinfo {author} {\bibfnamefont {J.~P.}\ \bibnamefont
  {Perdew}}, \bibinfo {author} {\bibfnamefont {K.}~\bibnamefont {Burke}}, \
  and\ \bibinfo {author} {\bibfnamefont {M.}~\bibnamefont {Ernzerhof}},\ }\href
  {\doibase 10.1103/PhysRevLett.77.3865} {\bibfield  {journal} {\bibinfo
  {journal} {Phys. Rev. Lett.}\ }\textbf {\bibinfo {volume} {77}},\ \bibinfo
  {pages} {3865} (\bibinfo {year} {1996})}\BibitemShut {NoStop}%
\bibitem [{\citenamefont {Rappe}\ \emph {et~al.}(1990)\citenamefont {Rappe},
  \citenamefont {Rabe}, \citenamefont {Kaxiras},\ and\ \citenamefont
  {Joannopoulos}}]{PhysRevB.41.1227}%
  \BibitemOpen
  \bibfield  {author} {\bibinfo {author} {\bibfnamefont {A.~M.}\ \bibnamefont
  {Rappe}}, \bibinfo {author} {\bibfnamefont {K.~M.}\ \bibnamefont {Rabe}},
  \bibinfo {author} {\bibfnamefont {E.}~\bibnamefont {Kaxiras}}, \ and\
  \bibinfo {author} {\bibfnamefont {J.~D.}\ \bibnamefont {Joannopoulos}},\
  }\href {\doibase 10.1103/PhysRevB.41.1227} {\bibfield  {journal} {\bibinfo
  {journal} {Phys. Rev. B}\ }\textbf {\bibinfo {volume} {41}},\ \bibinfo
  {pages} {1227} (\bibinfo {year} {1990})}\BibitemShut {NoStop}%
\end{thebibliography}
%

\end{document}